\documentclass[11pt]{article}
\usepackage[letterpaper,margin=1in]{geometry}
\usepackage[T1]{fontenc}
\usepackage{lmodern}
\usepackage{amsmath,amssymb,amsthm,mathtools}
\usepackage{booktabs,array}
\usepackage[square,numbers]{natbib}
\usepackage{tikz}
\usetikzlibrary{arrows.meta,positioning,calc}
\usepackage{enumitem}
\usepackage{needspace}
\usepackage[colorlinks=true,linkcolor=blue!50!black,citecolor=blue!50!black,urlcolor=blue!50!black]{hyperref}
\setlist{itemsep=2pt,topsep=4pt}
\allowdisplaybreaks[2]
\newtheorem{theorem}{Theorem}[section]

\theoremstyle{definition}
\newtheorem{definition}[theorem]{Definition}
\theoremstyle{remark}

\renewcommand{\binom}[2]{\genfrac{(}{)}{0pt}{0}{#1}{#2}}
\newcommand{\pfail}{p_{\mathrm{fail}}}
\newcommand{\Nbad}{N_{\mathrm{bad}}}
\newcommand{\BPM}{\operatorname{BPM}}

\title{Exact Failure Probabilities for $d$-Choice Cuckoo Hashing}
\author{
  Chenyang Zhang\\
  University of Illinois Urbana-Champaign\\
  \texttt{zcy331@illinois.edu}
  \and
  Ling Ren\\
  University of Illinois Urbana-Champaign\\
  \texttt{renling@illinois.edu}
}
\date{}
\begin{document}
\maketitle
\begin{abstract}
In $d$-choice cuckoo hashing, each of $n$ items must be placed in one of its $d$ candidate slots in a table of $M$ slots, and every slot holds at most one item. 
We give a polynomial-time dynamic programming algorithm to compute the exact probability that no valid placement exists. Our method also handles the variant with separate tables.

\end{abstract}
\section{Introduction}
\label{sec:introduction}

Cuckoo hashing has been used in a wide range of applications, including key-value stores~\citep{fan2013memc3}, network switches~\citep{zhou2013cuckooswitch}, approximate membership filters~\citep{fan2014cuckoo}, page tables~\citep{skarlatos2020elastic}, and cryptographic protocols~\citep{goodrich2011oram,pinkas2015phasing,angel2018pir}.

In the basic one-table variant of $d$-choice cuckoo hashing, each of $n$ items must be placed in one of its $d$ candidate slots in a table of $M$ slots, and every slot holds at most one item~\citep{pagh2004cuckoo,fotakis2005space}.
We study its failure probability, denoted $\pfail(n,M,d)$, that the candidate slots of the items admit no valid placement.

Previous work has established load thresholds for $d$-choice cuckoo
hashing~\citep{dietzfelbinger2010tight,frieze2012maximum,fountoulakis2012sharp},
and upper bounds on its failure
probability~\citep{fotakis2005space,yeo2023cuckoo, porat2012variant}.
The exact failure probability has also been obtained for $d=2$~\citep{kutzelnigg2006bipartite,kutzelnigg2008random,kutzelnigg2010further,drmota2012precise}, but has remained open for $d\ge3$~\citep[\S1.5.3]{kutzelnigg2008random}, \cite{angel2018pir}.

\paragraph{Our results.}
We compute the exact value of $\pfail(n,M,d)$ for any $n$, $M$, and $d$ in $O(n^2 \log n)$ time.
We use Beniamini's coefficient formula for perfect
matching~\citep{beniamini2022approximate} to express the number of failing
outcomes as a sum of exponentially many terms. We then compute this sum using dynamic programming.
Our method also 
extends to the variant with $d$ separate tables (possibly of unequal sizes), where it computes $\pfail(n, \mathrm m,d)$ in $O(n^{d+2})$ time. The implementations of our algorithms are available at \url{https://github.com/zcy05331/Cuckoo-Failure}.
\newpage

\section{Model and preliminaries}
\label{sec:model}

\subsection{Model}
\label{subsec:sampling}

There are $n$ items labeled $[n]=\{1,\ldots,n\}$.
In the \emph{one-table} variant, there are $M$ slots labeled
$[M]=\{1,\ldots,M\}$ with $1\le d<n<M$.
Each item $i$ independently draws a uniform $d$-element candidate set $\Gamma(i)\subseteq[M]$, and must be placed in one of the $d$ slots. Each item thus has
$Q=\binom Md$ possible candidate sets.

An \emph{outcome}
$\omega=(\Gamma(1),\ldots,\Gamma(n))$ lists the candidate sets of all
items, and the $Q^n$ outcomes are equally likely.
An outcome $\omega$ can be captured by a \emph{candidate graph} $G_\omega$:
a bipartite graph with the items on one side, the slots on the other,
and an edge $(i,j)$ for each $j\in\Gamma(i)$. A placement of the items is
a matching of $G_\omega$ that covers every item.
An outcome $\omega$ leads to a \emph{failure}
if $G_\omega$ has no such matching. Let $\Nbad$ be the number of failing outcomes. We have $\pfail=\Nbad/Q^n$.

We will later consider two
variants: independent draws with replacement
(Section~\ref{subsec:replacement}) and separate tables
(Section~\ref{sec:separate}).

\subsection{Beniamini's coefficient formula}
\label{subsec:beniamini}

\paragraph{Beniamini's expansion.}
Let $G$ be a bipartite graph with $M$ left vertices and $M$ right vertices, and let $K_{M,M}$ denote the complete bipartite graph on the same vertex set. For a
left vertex $i$ and a right vertex $j$, let $x_{ij}=1$ if the edge $(i,j)$
is missing from $G$ and $x_{ij}=0$ otherwise; we call $x=(x_{ij})_{i,j\in[M]}$ the
\emph{missing-edge indicators} of $G$.
Beniamini and Nisan~\citep{beniamini2021bipartite} analyze bipartite matching with a multilinear polynomial $\BPM_M^*(x)$.
$\BPM_M^*(x)$ is $1$ if $G$ has no perfect matching and $0$ otherwise. They show that this indicator polynomial has integer
coefficients as follows~\citep{beniamini2021bipartite,beniamini2022approximate}:
\begin{equation}
\BPM_M^*(x)=\sum_{H\subseteq K_{M,M}}a_H^*x^H,
\qquad x^H=\prod_{(i,j)\in E(H)}x_{ij}.
\label{eq:failure-polynomial}
\end{equation}
In the above, $H$ ranges over all bipartite graphs on the same $2M$ vertices.
Observe that $x^H=1$ if every edge of $H$ is missing from $G$ and 0 otherwise.
Beniamini~\citep{beniamini2022approximate} determined the coefficients $a_H^*$. To state his formula, we define totally ordered graphs.

\begin{definition}[Totally ordered graphs~\citep{beniamini2021bipartite,beniamini2022approximate}]
\label{def:chain}
Let $H$ be a bipartite graph and $T_H(i)=\{j:(i,j)\in E(H)\}$ be the
neighborhood of its $i$-th left vertex.
$H$ is \emph{totally ordered} if the
neighborhoods of its left vertices are totally ordered by inclusion: for all left vertices $i$ and $i'$, either $T_H(i)\subseteq T_H(i')$ or
$T_H(i')\subseteq T_H(i)$.
\end{definition}

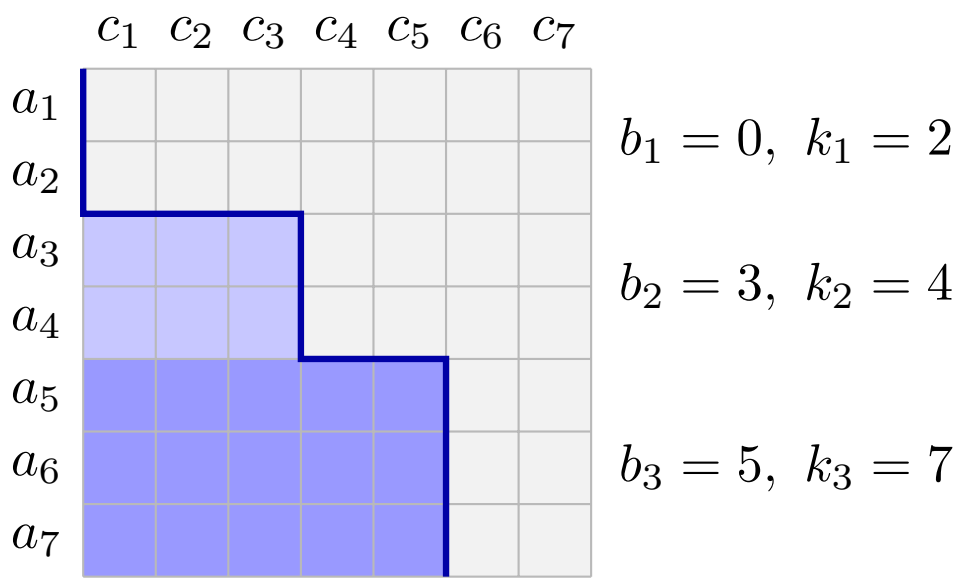
\begin{figure}[tbp]
\centering
\begin{tikzpicture}[x=0.48cm,y=0.48cm,>=Latex,font=\small]
\fill[gray!10] (0,0) rectangle (7,7);
\fill[blue!22] (0,3) rectangle (3,5);
\fill[blue!40] (0,0) rectangle (5,3);
\draw[step=1,gray!55,thin] (0,0) grid (7,7);
\draw[very thick,blue!65!black] (0,7)--(0,5)--(3,5)--(3,3)--(5,3)--(5,0);
\foreach \y/\txt in {6.5/1,5.5/2,4.5/3,3.5/4,2.5/5,1.5/6,0.5/7}
  \node[left] at (0,\y) {$a_{\txt}$};
\foreach \x in {1,...,7} \node[above] at (\x-0.5,7) {$c_{\x}$};
\node[right] at (7.1,6) {$b_1=0,\ k_1=2$};
\node[right] at (7.1,4) {$b_2=3,\ k_2=4$};
\node[right] at (7.1,1.5) {$b_3=5,\ k_3=7$};
\end{tikzpicture}
\caption{A totally ordered graph $H$ with $M=7$ vertices on each side.
Rows represent left vertices, columns represent right vertices, and a shaded cell
is an edge of $H$. The representing sequence is
$(k_i,b_i)=(2,0),(4,3),(7,5)$.
It indicates that two left vertices have no neighbors, two left vertices have (the same) three right neighbors, and three left vertices have (the same) five right neighbors.
}
\label{fig:staircase}
\end{figure}

For a totally ordered graph $H$, order the distinct neighborhoods of the left
vertices as $T_1\subsetneq T_2\subsetneq \cdots\subsetneq T_L$. The left vertices whose
neighborhood is $T_i$ form \emph{block}~$i$. Let $b_i=|T_i|$ be their
degree, and let $k_i$ be the number of left vertices in blocks
$1,\ldots,i$, with $k_0=0$:
\begin{equation}
0\le b_1<\cdots<b_L\le M,
\qquad 0=k_0<k_1<\cdots<k_L=M.
\label{eq:staircase}
\end{equation}
Block $i$ thus consists of $k_i-k_{i-1}$ left vertices, each of degree $b_i$.
$(\mathbf k,\mathbf b)=(k_i,b_i)_{i\le L}$ is called the \emph{representing sequence} of $H$. Figure~\ref{fig:staircase} gives an example. Throughout, we define $\binom ab=0$ unless $a,b$ are integers with $0\le b\le a$.
Empty products and $0^0$ equal one.

\begin{theorem}[{Beniamini~\citep[Theorem~2]{beniamini2022approximate}}]
\label{thm:beniamini}
If $H\subseteq K_{M,M}$ is not totally ordered, then $a_H^*=0$.
If $H$ is totally ordered with representing sequence~\eqref{eq:staircase},
then
\begin{equation}
a_H^*
=\binom{M-k_{L-1}-1}{M-b_L}
 \prod_{i=1}^{L-1}
 \phi(b_{i+1}-k_{i-1},b_i-k_{i-1},k_i-k_{i-1}),
\label{eq:chain-coefficient}
\end{equation}
where
\begin{equation}
\phi(a,b,r)=
\begin{cases}
\binom{a-1}{r},&b\le0,\\[2pt]
-\binom{a-b-1}{r-b}\binom{r-1}{b-1},&b>0.
\end{cases}
\label{eq:phi}
\end{equation}
\end{theorem}

The coefficient in~\eqref{eq:chain-coefficient} depends on $H$ only through
its representing sequence. We write it as $\mathfrak a(\mathbf k,\mathbf b)$.
Graphs that differ only by permutations of the two vertex sets have the
same coefficient.

\paragraph{Dummy items.}
A candidate graph $G_\omega$ has $n$ item vertices and $M$ slot vertices.
Since Beniamini's formula~\eqref{eq:failure-polynomial} requires an equal number of left vertices and right vertices, we add $M-n$ dummy items that can be placed in any slot. The resulting graph $\widehat G_\omega$ has $M$
vertices on each side, and it has a perfect matching if and only if
$G_\omega$ has a matching that covers every left vertex.


\section{Evaluation by dynamic programming}
\label{sec:quartic}

\subsection{Counting failing outcomes}
\label{subsec:expectation}

By~\eqref{eq:failure-polynomial}, the number of failing outcomes is
\begin{equation}
\Nbad=\sum_{\omega}\BPM_M^*(x(\omega))
=\sum_{H\subseteq K_{M,M}}a_H^*\sum_{\omega}x^H(\omega),
\label{eq:expectation-raw}
\end{equation}
where $\omega$ ranges over the $Q^n$ outcomes and $x(\omega)$ are the missing-edge indicators of $\widehat G_\omega$.

By Theorem~\ref{thm:beniamini}, only totally ordered graphs can have
$a_H^*\ne0$. If $H$ contains an edge incident to a dummy item, that edge
is present in $\widehat G_\omega$ for every outcome, so $x^H(\omega)=0$.
Thus, only totally ordered graphs in which every dummy item is isolated
contribute to~\eqref{eq:expectation-raw}.
Fix such a graph $H$. The monomial $x^H(\omega)$ equals $1$ if
and only if every item $i$ avoids its neighborhood $T_H(i)$, that
is, $\Gamma(i)\cap T_H(i)=\emptyset$. We call $T_H(i)$ the
\emph{forbidden set} of item $i$. If this set has size $b$, the number
of candidate sets that avoid it is
\begin{equation}
h(b)=\binom{M-b}{d}.
\label{eq:h}
\end{equation}
The candidate sets of the $n$ real items can be chosen independently, so
\begin{equation}
\sum_{\omega}x^H(\omega)=\prod_{i=1}^{n}h\bigl(|T_H(i)|\bigr).
\label{eq:monomial-count}
\end{equation}
Substituting into~\eqref{eq:expectation-raw} gives
\begin{equation}
\Nbad=\sum_{H}
a_H^*\prod_{i=1}^{n}h\bigl(|T_H(i)|\bigr),
\label{eq:weighted-chain-sum}
\end{equation}
where $H$ ranges over the totally ordered graphs in which every dummy
item is isolated.
\paragraph{Grouping by representing sequences.}
Consider a graph $H$ that contributes to~\eqref{eq:weighted-chain-sum}.
Let $(\mathbf k,\mathbf b)$ be its representing sequence. Its dummy
items are isolated, so they have degree $0$ and lie in the first block.
Hence $b_1=0$ and $k_1\ge z$, where $z=M-n>0$ is the number of dummy items.

Let $u_1=k_1-z$ and $u_i=k_i-k_{i-1}$ for $2\le i\le L$.
Then $u_i$ is the number of real items in block $i$, and $\sum_i u_i=n$.
The term associated with $H$ in~\eqref{eq:weighted-chain-sum} can be rewritten as
\[
\mathfrak a(\mathbf k,\mathbf b)\prod_i h(b_i)^{u_i}
\]
because every item
of block $i$ has a forbidden set of size $b_i$.

The number of graphs with a given
representing sequence can be computed as follows. The $n$ real items are
distributed among the blocks in $n!/\prod_iu_i!$ ways. The nested sets
$T_1\subsetneq\cdots\subsetneq T_L$ are chosen in
$\binom{M}{b_1} \prod_{i=1}^{L-1}\binom{M-b_i}{b_{i+1}-b_i}$ ways. Since $b_1= 0$, the factor $\binom{M}{b_1}=1$ can be omitted. 
Let $\mathcal S_L(n,M)$ be the set of representing sequences with $L$ blocks with $b_1=0$ and $k_1\ge z$. We have
\begin{equation}
\Nbad(n,M,d)
=\sum_{L=1}^{M}\ \sum_{(\mathbf k,\mathbf b)\in\mathcal S_L(n,M)}
\mathfrak a(\mathbf k,\mathbf b)\frac{n!}{\prod_{i=1}^L u_i!}
\prod_{i=1}^{L-1}\binom{M-b_i}{b_{i+1}-b_i}
\prod_{i=1}^L h(b_i)^{u_i}
\label{eq:explicit-sum}
\end{equation}

Direct evaluation would have to sum exponentially many nonzero representing sequences.
We compute this sum with dynamic programming.

\subsection{The forward recurrence}
\label{subsec:forward-state}

\paragraph{Blockwise factorization of each summand.}
For each representing sequence, we first express its contribution to~\eqref{eq:explicit-sum} as a product of blockwise factors.
The coefficient $\mathfrak a(\mathbf k,\mathbf b)$ already has this form according to~\eqref{eq:chain-coefficient}. Each block
$i<L$ contributes $\phi(b_{i+1}-k_{i-1},b_i-k_{i-1},k_i-k_{i-1})$, and the
last block contributes $\binom{M-k_{L-1}-1}{M-b_L}$. The
$\binom{M-b_i}{b_{i+1}-b_i}$ term belongs to block $i$ by definition. So does the $h(b_i)^{u_i}$ term. The multinomial coefficient can also be decomposed to individual blocks:
\begin{equation}
\frac{n!}{\prod_{i=1}^{L}u_i!}
=\prod_{i=1}^{L}\binom{n-\sum_{j<i}u_j}{u_i},
\label{eq:multinomial-split}
\end{equation}
where $\binom{n-\sum_{j<i}u_j}{u_i}$ chooses the $u_i$ items of block $i$ among the items not yet placed.

For a nonfinal block $i<L$, we can think of $(k_{i-1},b_i)$ as the
\emph{boundary} at its start and $(k_i,b_{i+1})$ as the boundary at its end.
Let $\nu(\ell)=\max\{0,\ell-z\}$ be the number of real items among the
first $\ell$ left vertices. The factors associated with a nonfinal block with
boundaries $(\ell,b)$ and $(r,c)$ are
\begin{equation}
\tau(\ell,b;r,c)
=\phi(c-\ell,b-\ell,r-\ell)
\binom{M-b}{c-b}
\binom{n-\nu(\ell)}{u}
h(b)^{u},
\qquad u=r-\max\{\ell,z\}.
\label{eq:block-multiplier}
\end{equation}
For a fixed representing sequence, the corresponding summand in~\eqref{eq:explicit-sum} can now be written as
\begin{equation}
\prod_{i=1}^{L-1}\tau(k_{i-1},b_i;k_i,b_{i+1})
\cdot\binom{M-k_{L-1}-1}{M-b_L}h(b_L)^{u_L},
\label{eq:term-as-product}
\end{equation}
which consists of one multiplier per nonfinal block, and a factor for the final block.

\paragraph{Prefixes with the same boundary can be merged.}
The factors in~\eqref{eq:block-multiplier} depend only on the block's
start and end boundaries, not on earlier blocks.
Consider the first $i<L$ blocks of a representing sequence.
We define a \emph{prefix} to consist of these
$i$ blocks together with $b_{i+1}$. 
(Note that the end boundary of block $i$ is $b_{i+1}$.)
The \emph{weight} of the prefix is
\[
\prod_{j=1}^{i}
\tau(k_{j-1},b_j;k_j,b_{j+1}).
\]

Now extend a prefix by a nonfinal block with start boundary $(\ell, b)$ and end boundary $(r, c)$. Recall that each representing sequence corresponds to a single summand in \eqref{eq:explicit-sum}, which \eqref{eq:term-as-product} factors into blockwise multipliers. Thus, for any complete representing sequence extending the prefix, its contribution to \eqref{eq:explicit-sum} factors via \eqref{eq:term-as-product} as

\[
(\text{weight of the prefix})
\times\tau(\ell,b;r,c)
\times(\text{factors of the later blocks}).
\]
The multiplier $\tau(\ell,b;r,c)$ is the same for every prefix ending at
$(\ell,b)$. We can therefore sum the weights of these prefixes before
extending them. 

\paragraph{States and recurrence.}
Define the \emph{forward state} $F[\ell,b]$ as the sum
of the weights of all prefixes ending at $(\ell,b)$.
The empty prefix has the empty forbidden set and is initialized to
\begin{equation}
F[0,0]=1.
\label{eq:forward-init}
\end{equation}
All other entries are initially zero. By the binomial convention,
$\tau(\ell,b;r,c)$ in~\eqref{eq:block-multiplier} vanishes unless $z\le r$
and $c>r$, so only such entries are assigned below. Extending the prefixes that end at each
boundary $(\ell,b)$ to $(r,c)$ gives, for $z\le r<M$ and
$r<c\le M$,
\begin{equation}
F[r,c]=\sum_{\ell=0}^{r-1}\sum_{b=0}^{c-1}F[\ell,b]\,\tau(\ell,b;r,c).
\label{eq:forward-transition}
\end{equation}
A final block takes the terminal
factor of~\eqref{eq:term-as-product} with all remaining items, so
\begin{equation}
\Nbad=\sum_{\ell=0}^{M-1}\sum_{b=0}^{M}
F[\ell,b]\binom{M-\ell-1}{M-b}h(b)^{n-\nu(\ell)}.
\label{eq:forward-terminal}
\end{equation}

Each summand $F[\ell,b]\,\tau(\ell,b;r,c)$ of~\eqref{eq:forward-transition}
extends the prefixes ending at $(\ell,b)$ by one block. 
By induction on $\ell$, the recurrence computes
$F[\ell,b]$ as defined, and~\eqref{eq:forward-terminal} equals the
sum~\eqref{eq:explicit-sum}. 

With $O(M^2)$ states and $O(M^2)$ predecessors per state, the recurrence uses $O(M^4)$ integer
operations and $O(M^2)$ cells. Its states depend on $M$; the next
subsection removes this dependence.


\subsection{Faster evaluation and reuse across table sizes}
\label{sec:cubic}
\label{subsec:coordinates}

We further improve the algorithm to \(O(n^2\log n)\) by combining three observations.

First, after the initial block of degree 0, the total weight of the remaining blocks depends only on the numbers of remaining real items \(p = M - \ell\) and available slots \(v = M - b\), where \(0\le v<p\le n\). We call the remaining blocks a \emph{suffix} and let \(V[p,v]\) denote their total weight. This gives \(O(n^2)\) states independent of the table size \(M\).

Second, in the recurrence for \(V[p,v]\), the inner sum over the next available-slot count \(w\) depends only on \(v\) and the next item count \(q\), but not on \(p\). Reusing these shared sums splits the transition into two stages and reduces the complexity from \(O(n^4)\) to \(O(n^3)\).

Third, the binomial coefficients in both stages can be rearranged into convolution kernels, allowing both the shared sums and the suffix weights to be computed using polynomial multiplication. 
Using FFT-based polynomial multiplication, the complexity is reduced to \(O(n^2\log n)\).


\subsection{Independent draws with replacement}
\label{subsec:replacement}

Suppose instead that each item makes $d$ independent uniform draws from $[M]$.
Its sampling outcome is an ordered \(d\)-tuple, so $Q=M^d$. Of these, $h(b)=(M-b)^d$ avoid a forbidden set of size $b$. Apart from these two counts, the algorithms in this section stay unchanged.


\section{Evaluation for separate tables}
\label{sec:separate}

\subsection{The model and the state}
\label{subsec:separate-model}

There are $d$ disjoint tables of positive lengths $m_1,\ldots,m_d$. Each
item chooses one slot uniformly from each table, independently across
tables and items, and must occupy one of these slots. Write
$\mathbf m=(m_1,\ldots,m_d)$, $M=\sum_jm_j$, and $Q=\prod_jm_j$. There
are $Q^n$ equally likely outcomes; let $\Nbad(n;\mathbf m)$
count those admitting no valid placement. We consider $d<n<M$.

The recurrence in Section~\ref{subsec:forward-state} tracks the total number of forbidden slots $b$. For separate tables, we need to track the number of forbidden slots in each table. For example, with two tables of length $4$, the forbidden counts $(3,1)$ and $(2,2)$ both total four but
leave $1\cdot3=3$ and $2\cdot2=4$ candidate choices per item, respectively. Let
$\mathbf b=(b_1,\ldots,b_d)$ denote the number of forbidden slots in each table, and write
$|\mathbf b|=\sum_jb_j$. An item whose forbidden set has counts
$\mathbf b$ has
\begin{equation}
h_{\mathbf m}(\mathbf b)=\prod_{j=1}^d(m_j-b_j)
\label{eq:split-h}
\end{equation}
candidate choices. 

For two consecutive blocks, let \(\mathbf b\) and \(\mathbf c=(c_1,\ldots,c_d)\) denote the counts of their respective forbidden sets in each table. The latter is obtained from the former by choosing \(c_j-b_j\) additional slots from the \(m_j-b_j\) remaining slots in each table \(j\). The number of possible choices is

\begin{equation}
\binom{\mathbf m-\mathbf b}{\mathbf c-\mathbf b}
=\prod_{j=1}^d\binom{m_j-b_j}{c_j-b_j}.
\label{eq:split-binomial}
\end{equation}
This binomial notation will also be used for other quantities.

\subsection{The forward recurrence}
\label{subsec:separate-forward}

Define the \emph{forward state} $F_d[\ell,\mathbf b]$ as in
Section~\ref{subsec:forward-state}: the sum of the weights of all
prefixes that end at left vertex $\ell$ with a forbidden set of counts
$\mathbf b$.

The multiplier~\eqref{eq:block-multiplier}
changes in two places: the candidate count $h(b)$ becomes
$h_{\mathbf m}(\mathbf b)$, and the set extension $\binom{M-b}{c-b}$
becomes $\binom{\mathbf m-\mathbf b}{\mathbf c-\mathbf b}$. The
coefficient factor and the item-choice factor are unchanged:
\begin{equation}
\tau_d(\ell,\mathbf b;r,\mathbf c)
=\phi(|\mathbf c|-\ell,|\mathbf b|-\ell,r-\ell)
\binom{\mathbf m-\mathbf b}{\mathbf c-\mathbf b}
\binom{n-\nu(\ell)}{u}
h_{\mathbf m}(\mathbf b)^{u},
\label{eq:split-multiplier}
\end{equation}
with $u=r-\max\{\ell,z\}$, $z=M-n$ dummy items, and
$\nu(\ell)=\max\{0,\ell-z\}$ as before.

Initialize $F_d[0,\mathbf0]=1$ and all other entries to zero. For
$z\le r<M$ and $\mathbf0\le\mathbf c\le\mathbf m$ with $|\mathbf c|>r$,
\begin{equation}
F_d[r,\mathbf c]
=\sum_{\ell=0}^{r-1}\ \sum_{\mathbf0\le\mathbf b<\mathbf c}
F_d[\ell,\mathbf b]\,\tau_d(\ell,\mathbf b;r,\mathbf c).
\label{eq:split-forward}
\end{equation}
A final block takes all remaining items, so
\begin{equation}
\Nbad(n;\mathbf m)
=\sum_{\ell=0}^{M-1}\ \sum_{\mathbf0\le\mathbf b\le\mathbf m}
F_d[\ell,\mathbf b]\binom{M-\ell-1}{M-|\mathbf b|}
h_{\mathbf m}(\mathbf b)^{n-\nu(\ell)},
\label{eq:split-terminal}
\end{equation}
and the failure probability is $\Nbad(n;\mathbf m)/Q^n$.


\subsection{Complexity}
\label{subsec:separate-complexity}

Apart from $(0,\mathbf0)$, the recurrence~\eqref{eq:split-forward}
produces states with $z\le\ell<M$ and $|\mathbf b|>\ell$ only. If
$b_j=m_j$ for some $j$, then $h_{\mathbf m}(\mathbf b)=0$: every later
item lacks a candidate in table $j$, so every transition from the state
and its term in~\eqref{eq:split-terminal} vanish. We therefore keep only
states with $b_j<m_j$ for every $j$. Put $p=M-\ell$ and
$\mathbf v=\mathbf m-\mathbf b$ for the remaining items and the available
slots in each table. The retained states satisfy
\[
1\le p\le n,\qquad \mathbf v\ge\mathbf1,\qquad |\mathbf v|<p.
\]
For each $p$ there are at most $\binom{p-1}{d}$ such vectors $\mathbf v$,
so the recurrence has at most
\[
1+\sum_{p=1}^{n}\binom{p-1}{d}=1+\binom{n}{d+1}=O(n^{d+1})
\]
states.
Each state is computed by enumerating all retained predecessors and
applying~\eqref{eq:split-multiplier}. With the binomial coefficients and
the powers of $h_{\mathbf m}$ precomputed, a transition costs $O(1)$
operations. Hence the recurrence
uses $O(n^{2d+2})$ integer operations and $O(n^{d+1})$ cells for
each table-length vector $\mathbf m$. This is polynomial in $n$ for
every constant $d$.

\paragraph{Faster evaluation.}
The suffix formulation of Section~\ref{sec:cubic} also applies to
separate tables. Tracking suffixes by the number of remaining items and
the available counts in each table gives a table that depends only on
$n$ and $d$. Sharing inner sums and applying two binomial transforms
builds it in $O(n^{d+2})$ operations, after which each table-length
vector is evaluated in $O(n^{d+1})$ operations.

\section{Conclusion}
\label{sec:conclusion}

We present dynamic programming algorithms for the exact failure probability of $d$-choice cuckoo hashing for both the one-table and separate-table variants. The algorithms evaluate
Beniamini's expansion by combining prefixes with the same block
boundary. For the one-table variant, our algorithm runs in $O(n^2 \log n)$ time.
For the separate-table variant, our algorithm runs in $O(n^{d+2})$ time for constant $d$.
A direction for future work is to find an algorithm for separate tables that is polynomial in both $n$ and $d$.


\label{main:end}
\begingroup
\small
\setlength{\bibsep}{0pt}
\bibliographystyle{alpha}
\bibliography{references}

@inproceedings{beniamini2022approximate,
author = {Beniamini, Gal},
title = {The approximate degree of bipartite perfect matching},
year = {2022},
isbn = {9783959772419},
publisher = {Schloss Dagstuhl--Leibniz-Zentrum fuer Informatik},
address = {Dagstuhl, DEU},
url = {https://doi.org/10.4230/LIPIcs.CCC.2022.1},
doi = {10.4230/LIPIcs.CCC.2022.1},
booktitle = {Proceedings of the 37th Computational Complexity Conference},
articleno = {1},
numpages = {26},
location = {Philadelphia, PA, USA},
series = {CCC '22}
}

@inproceedings{beniamini2021bipartite,
  title={Bipartite perfect matching as a real polynomial},
  author={Beniamini, Gal and Nisan, Noam},
  booktitle={Proceedings of the 53rd Annual ACM SIGACT Symposium on Theory of Computing},
  pages={1118--1131},
  year={2021}
}

@inproceedings{kutzelnigg2006bipartite,
  author    = {Reinhard Kutzelnigg},
  title     = {Bipartite Random Graphs and Cuckoo Hashing},
  booktitle = {Fourth Colloquium on Mathematics and Computer Science},
  series    = {DMTCS Proceedings},
  volume    = {AG},
  pages     = {403--406},
  year      = {2006},
  doi       = {10.46298/dmtcs.3486},
  url       = {https://dmtcs.episciences.org/3486/pdf}
}

@article{kutzelnigg2010further,
  author    = {Reinhard Kutzelnigg},
  title     = {A Further Analysis of Cuckoo Hashing with a Stash and Random Graphs of Excess {$r$}},
  journal   = {Discrete Mathematics \& Theoretical Computer Science},
  volume    = {12},
  number    = {3},
  pages     = {81--102},
  year      = {2010},
  doi       = {10.46298/dmtcs.495},
  url       = {https://dmtcs.episciences.org/495/pdf}
}

@inproceedings{dietzfelbinger2010tight,
  title={Tight thresholds for cuckoo hashing via XORSAT},
  author={Dietzfelbinger, Martin and Goerdt, Andreas and Mitzenmacher, Michael and Montanari, Andrea and Pagh, Rasmus and Rink, Michael},
  booktitle={International Colloquium on Automata, Languages, and Programming},
  pages={213--225},
  year={2010},
  organization={Springer}
}

@inproceedings{yeo2023cuckoo,
  author    = {Kevin Yeo},
  title     = {Cuckoo Hashing in Cryptography: Optimal Parameters, Robustness and Applications},
  booktitle = {Advances in Cryptology -- CRYPTO 2023},
  pages     = {197--230},
  publisher = {Springer Nature Switzerland},
  year      = {2023},
  doi       = {10.1007/978-3-031-38551-3_7},
  url       = {https://doi.org/10.1007/978-3-031-38551-3_7},
  eprint    = {2306.11220},
  archivePrefix = {arXiv}
}

@article{pagh2004cuckoo,
  author    = {Rasmus Pagh and Flemming Friche Rodler},
  title     = {Cuckoo Hashing},
  journal   = {Journal of Algorithms},
  volume    = {51},
  number    = {2},
  pages     = {122--144},
  year      = {2004},
  doi       = {10.1016/j.jalgor.2003.12.002},
  url       = {https://doi.org/10.1016/j.jalgor.2003.12.002}
}

@article{fotakis2005space,
  author    = {Dimitris Fotakis and Rasmus Pagh and Peter Sanders and Paul Spirakis},
  title     = {Space Efficient Hash Tables with Worst Case Constant Access Time},
  journal   = {Theory of Computing Systems},
  volume    = {38},
  number    = {2},
  pages     = {229--248},
  year      = {2005},
  doi       = {10.1007/s00224-004-1195-x},
  url       = {https://doi.org/10.1007/s00224-004-1195-x}
}

@article{frieze2012maximum,
  author    = {Alan Frieze and P{\'a}ll Melsted},
  title     = {Maximum Matchings in Random Bipartite Graphs and the Space Utilization of Cuckoo Hash Tables},
  journal   = {Random Structures \& Algorithms},
  volume    = {41},
  number    = {3},
  pages     = {334--364},
  year      = {2012},
  doi       = {10.1002/rsa.20427},
  url       = {https://doi.org/10.1002/rsa.20427}
}

@article{fountoulakis2012sharp,
  author    = {Nikolaos Fountoulakis and Konstantinos Panagiotou},
  title     = {Sharp Load Thresholds for Cuckoo Hashing},
  journal   = {Random Structures \& Algorithms},
  volume    = {41},
  number    = {3},
  pages     = {306--333},
  year      = {2012},
  doi       = {10.1002/rsa.20426},
  url       = {https://doi.org/10.1002/rsa.20426}
}

@inproceedings{porat2012variant,
  title={A cuckoo hashing variant with improved memory utilization and insertion time},
  author={Porat, Ely and Shalem, Bar},
  booktitle={2012 Data Compression Conference},
  pages={347--356},
  year={2012},
  organization={IEEE}
}

@article{drmota2012precise,
  author    = {Michael Drmota and Reinhard Kutzelnigg},
  title     = {A Precise Analysis of Cuckoo Hashing},
  journal   = {ACM Transactions on Algorithms},
  volume    = {8},
  number    = {2},
  pages     = {11:1--11:36},
  year      = {2012},
  doi       = {10.1145/2151171.2151174},
  url       = {https://doi.org/10.1145/2151171.2151174}
}

@inproceedings{angel2018pir,
  author    = {Sebastian Angel and Hao Chen and Kim Laine and Srinath Setty},
  title     = {{PIR} with Compressed Queries and Amortized Query Processing},
  booktitle = {2018 IEEE Symposium on Security and Privacy (SP)},
  pages     = {962--979},
  publisher = {IEEE},
  year      = {2018},
  doi       = {10.1109/SP.2018.00062},
  url       = {https://doi.org/10.1109/SP.2018.00062}
}

@phdthesis{kutzelnigg2008random,
  author    = {Reinhard Kutzelnigg},
  title     = {Random Bipartite Graphs and their Application to Cuckoo Hashing},
  school    = {Technische Universit{\"a}t Wien},
  year      = {2008},
  url       = {https://repositum.tuwien.at/handle/20.500.12708/14415}
}

@inproceedings{goodrich2011oram,
  author    = {Michael T. Goodrich and Michael Mitzenmacher},
  title     = {Privacy-Preserving Access of Outsourced Data via Oblivious {RAM} Simulation},
  booktitle={International Colloquium on Automata, Languages, and Programming},
  pages={576--587},
  year={2011},
  organization={Springer}
}

@inproceedings{pinkas2015phasing,
  author    = {Benny Pinkas and Thomas Schneider and Gil Segev and Michael Zohner},
  title     = {Phasing: Private Set Intersection Using Permutation-based Hashing},
  booktitle = {24th {USENIX} Security Symposium},
  pages     = {515--530},
  publisher = {{USENIX} Association},
  year      = {2015}
}

@inproceedings{fan2013memc3,
  author    = {Bin Fan and David G. Andersen and Michael Kaminsky},
  title     = {{MemC3}: Compact and Concurrent {MemCache} with Dumber Caching and Smarter Hashing},
  booktitle = {10th {USENIX} Symposium on Networked Systems Design and Implementation ({NSDI} 13)},
  pages     = {371--384},
  publisher = {{USENIX} Association},
  year      = {2013},
  url       = {https://www.usenix.org/conference/nsdi13/technical-sessions/presentation/fan}
}

@inproceedings{zhou2013cuckooswitch,
  title={Scalable, high performance ethernet forwarding with cuckooswitch},
  author={Zhou, Dong and Fan, Bin and Lim, Hyeontaek and Kaminsky, Michael and Andersen, David G},
  booktitle={Proceedings of the ninth ACM conference on Emerging networking experiments and technologies},
  pages={97--108},
  year={2013}
}

@inproceedings{fan2014cuckoo,
  author    = {Bin Fan and David G. Andersen and Michael Kaminsky and Michael D. Mitzenmacher},
  title     = {Cuckoo Filter: Practically Better Than {Bloom}},
  booktitle={Proceedings of the 10th ACM International on Conference on emerging Networking Experiments and Technologies},
  pages     = {75--88},
  publisher = {ACM},
  year      = {2014},
  doi       = {10.1145/2674005.2674994},
  url       = {https://doi.org/10.1145/2674005.2674994}
}

@inproceedings{skarlatos2020elastic,
  title={Elastic cuckoo page tables: Rethinking virtual memory translation for parallelism},
  author={Skarlatos, Dimitrios and Kokolis, Apostolos and Xu, Tianyin and Torrellas, Josep},
  booktitle={Proceedings of the Twenty-Fifth International Conference on Architectural Support for Programming Languages and Operating Systems},
  pages={1093--1108},
  year={2020}
}
\endgroup
\end{document}